\documentclass{main}

\usepackage[backend=biber,style=numeric-comp,natbib=true,sorting=none,maxnames=3]{biblatex}
\usepackage{caption}
\usepackage{placeins}
\usepackage{subcaption}

\def\snn{\sqrt{s_{\mathrm{NN}}}}
\def\jpsi{\ensuremath{\mathrm{J}\kern-0.02em/\kern-0.05em\psi}}
\def\pt{p_{\rm T}}

\def\be{\begin{equation}}
\def\ee{\end{equation}}
\def\bea{\begin{eqnarray}}
\def\eea{\end{eqnarray}}

\newcommand{\Photo}{\includegraphics[height=35mm]{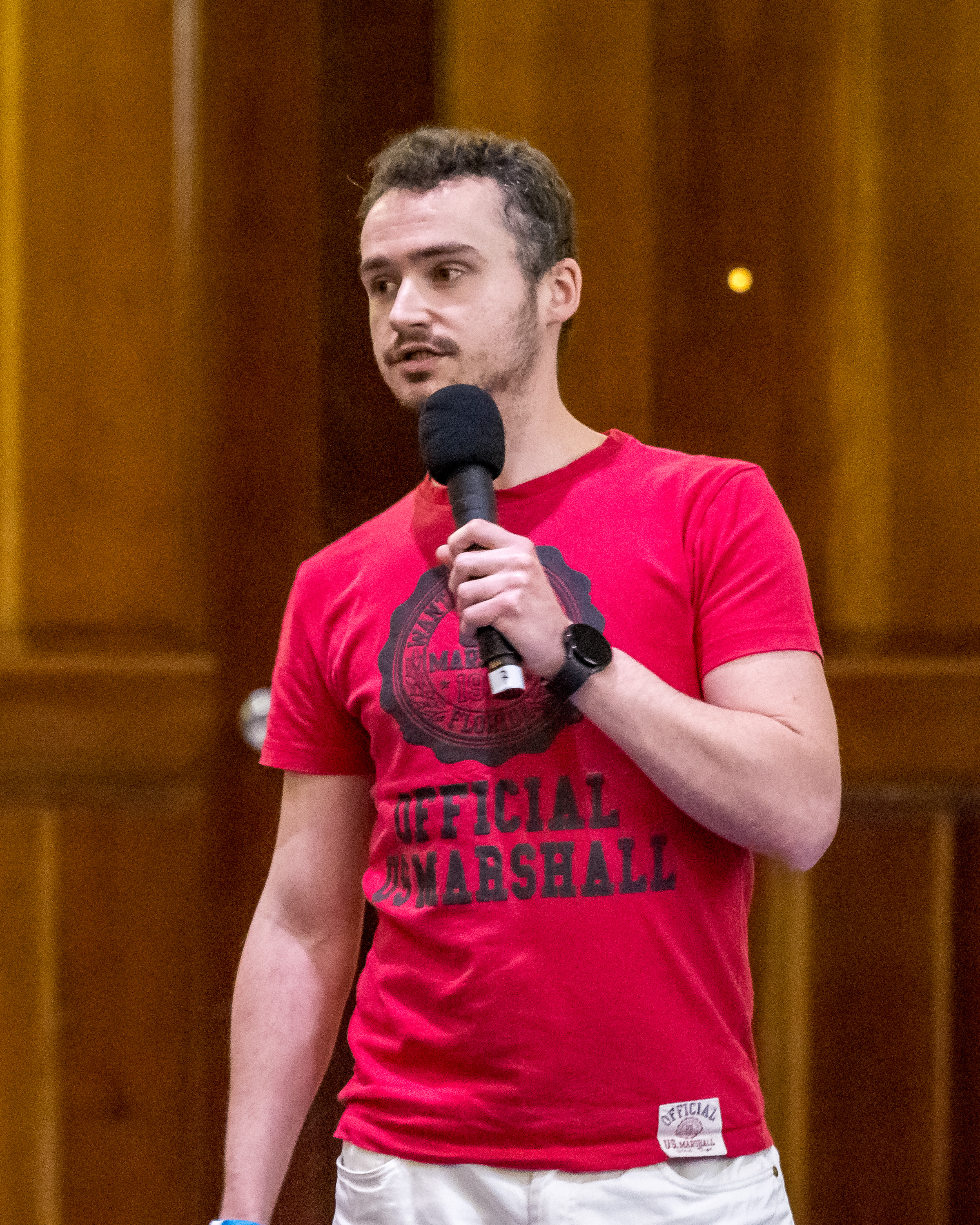}}
\begin{document}

\title{\Large Photoproduction of J/$\psi$ and dileptons in Pb--Pb collisions with nuclear overlap}

\author{N. Bizé (for the ALICE Collaboration)}

\address{SUBATECH, Nantes Université, IMT-Atlantique, CNRS.\\ Nantes, France
}

\maketitle\abstracts{
Photon-photon reactions and the production of J/$\psi$ meson through photonuclear reactions have been extensively studied in ultra-peripheral heavy-ion collisions, in which the impact parameter is larger than twice the nuclear radius. In recent years, coherently photoproduced J/$\psi$ and dilepton production via photon-photon interactions have also been observed in nucleus-nucleus (A--A) collisions with nuclear overlap. The former can help to constrain the nuclear gluon distributions at low Bjorken-$x$ and high energy, while the latter could be used to further map the electromagnetic fields produced in heavy-ion collisions. In addition, these measurements can shed light on the theory behind photon-induced reactions in A--A collisions with nuclear overlap, including possible interactions of the measured probes with the formed and fast expanding quark-gluon plasma.
Since the produced quarkonium is expected to keep the polarization of the incoming photon due to $s$-channel helicity conservation, the photoproduction origin of the $\jpsi$ yield excess at very low transverse momentum, $\pt$, can be confirmed by the measurement of the $\jpsi$ polarization.
The ALICE detector can perform quarkonium production measurements at both mid ($|y|<0.9$) and forward ($2.5<y<4$) rapidities down to $p_{\rm T} = 0$. In the following, the new ALICE measurements of the $\jpsi$ $y$-differential cross section and the first polarization results of coherently photoproduced $\jpsi$ via the dimuon decay channel at forward rapidity in Pb--Pb collisions at $\snn=$ 5.02 TeV are reported. Additionally, the measurement of an excess with respect to expectations from hadronic production in the dielectron yield, at low mass and $p_{\rm T}$, at midrapidity in Pb--Pb collisions at $\snn=$ 5.02~TeV, is presented. The results are compared with available theoretical models.
}

\section{Introduction}

Strong electromagnetic fields are generated in ultra-peripheral collisions (UPC), where the impact parameter is larger than the sum of the radii of the two colliding nucleus. In such collisions, hadronic interactions are strongly suppressed. UPC provide therefore a clean environment to study vector meson and dilepton photoproduction and extensive measurements have been carried out. 
Recently, the aforementioned probes have also been studied in events with nuclear overlap, where hadronic interactions become dominant.
In such events, the vector meson photoproduction can help to probe the gluon distribution in the nuclei for different Bjorken-$x$ regions. In addition, it allows to test the survival of the coherence process while the nuclei are broken during the hadronic collision. The dilepton photoproduction can provide constraints on the mapping of the electromagnetic fields that are generated by the fast moving charges around the nuclei. Both processes could be useful to look for possible final-state medium effects.
Previous measurements of dielectron photoproduction with nuclear overlap have been performed at STAR~\cite{PhysRevLett.121.132301} and ATLAS~\cite{PhysRevLett.121.212301}, including an excess found at low $\pt$ with respect to expectations from known hadronic sources.
Furthermore, a $\jpsi$ yield excess has been observed at very low $\pt$, for $\pt <$ 0.3~GeV/$c$ and forward rapidity in peripheral Pb--Pb collisions at $\snn = $~2.76 TeV~\cite{ALICE:2015mzu} and 5.02~TeV~\cite{ALICE:2022zso}. The coherently photoproduced $\jpsi$ cross section as a function of centrality in this $\pt$ region has been measured in Pb--Pb collisions with nuclear overlap~\cite{ALICE:2022zso}. The theoretical models~\cite{GayDucati:2018who} predict a strong rapidity dependence of the vector meson cross section, in particular at forward rapidity.
Finally, previous measurements in UPC~\cite{ALICE:2023svb} indicate a transverse polarization of the coherently photoproduced $\jpsi$. While the latter is expected to keep the polarization of the incoming photon due to s-channel helicity conservation, such measurement in events with nuclear overlap is important to confirm the photoproduction origin of the very low-$\pt$ $\jpsi$ excess.
In the following, new ALICE measurements of the $y$--differential cross section of the coherently photoproduced $\jpsi$ and the inclusive $\jpsi$ polarization at forward rapidity in Pb--Pb collisions at $\snn =$~5.02~TeV with nuclear overlap are presented. In addition, an excess of the photoproduced dielectron yield with respect to the expectations from the hadronic production, at low mass and $\pt$, at midrapidity in Pb--Pb collisions at $\snn =$ 5.02~TeV, is reported.

\section{Experimental apparatus}

A detailed description of the ALICE experiment can be found in~\cite{ALICE:2008ngc}. In the following results, dielectron measurements are performed at midrapidity ($|y|<$ 0.9) with the ALICE central barrel. The main detectors used for tracking and electron identification are the Inner Tracking System (ITS), the Time Projection Chamber (TPC) and the Time-Of-Flight (TOF) system.
All three detectors cover at least the pseudorapidity acceptance $|\eta| < 0.9$.
The ITS is composed of six concentric cylindrical layers of silicon detectors. It is designed to locate the primary collision vertex with a spatial resolution of the order of 100 $\mu$m. In addition, it allows tracking and particle identification (PID) for very low $\pt$, $\pt$ $<$ 0.2 GeV/$c$.
The TPC is a cylindrical detector filled with a gas mixture (Ne/CO$_2$/N$_2$) containing a central cathode and two end plates each composed of 18 Multiple Wired Proportional Chambers (MWPC). The primary electrons drift into the gas mixture on either side of the cathode to the end plates. TPC is used for PID by measuring specific energy loss $\mathrm{d}E/\mathrm{d}x$, tracking and vertex reconstruction in a large $\pt$ interval ranging from 0.1 to 100 GeV/$c$. 
The TOF is a cylindrical detector composed of Multiple Resistive Plate Chambers (MRPCs). The use of MRPCs technology allows one to reach the required time resolution of less than 100 ps to identify particles using their time-of-flight for $\pt<2.5$ GeV/$c$.
The $\jpsi$ is studied via the dimuon decay channel at forward rapidity (2.5 $< y <$ 4) with the ALICE muon spectrometer.
The latter is composed of a tracking system, the Muon Chambers (MCH) and a trigger system, the Muon Trigger (MTR). 
MCH is made of five tracking stations, each one formed of two chambers based on the technology of the MWPCs. The detector is designed to reach a spatial resolution of the order of 100 $\mu$m to obtain the required invariant mass resolution for quarkonium analysis ($\sim$ 100 MeV/$c^2$ for the $\Upsilon$). 
MTR is composed of four Resistive Plate Chambers (RPCs), which are allocated in two stations. The purpose of this trigger system is to reduce the probability to keep an event containing only muons coming from $\pi$ or K. It is performed by applying a configurable $\pt$--threshold ranging from 0.5 to 4.2 GeV/$c$.
A front absorber is placed upstream from MCH to suppress the background coming from hadrons and secondary muons.
An iron wall is installed between MCH and MTR to filter out the muons originating from the remaining hadrons while a rear absorber is placed downstream from MTR to protect the spectrometer from beam-background interactions.

\section{Results}

\subsection{Dielectron photoproduction measurement at midrapidity}

\begin{figure}
    \centering
    \includegraphics[scale=0.3]{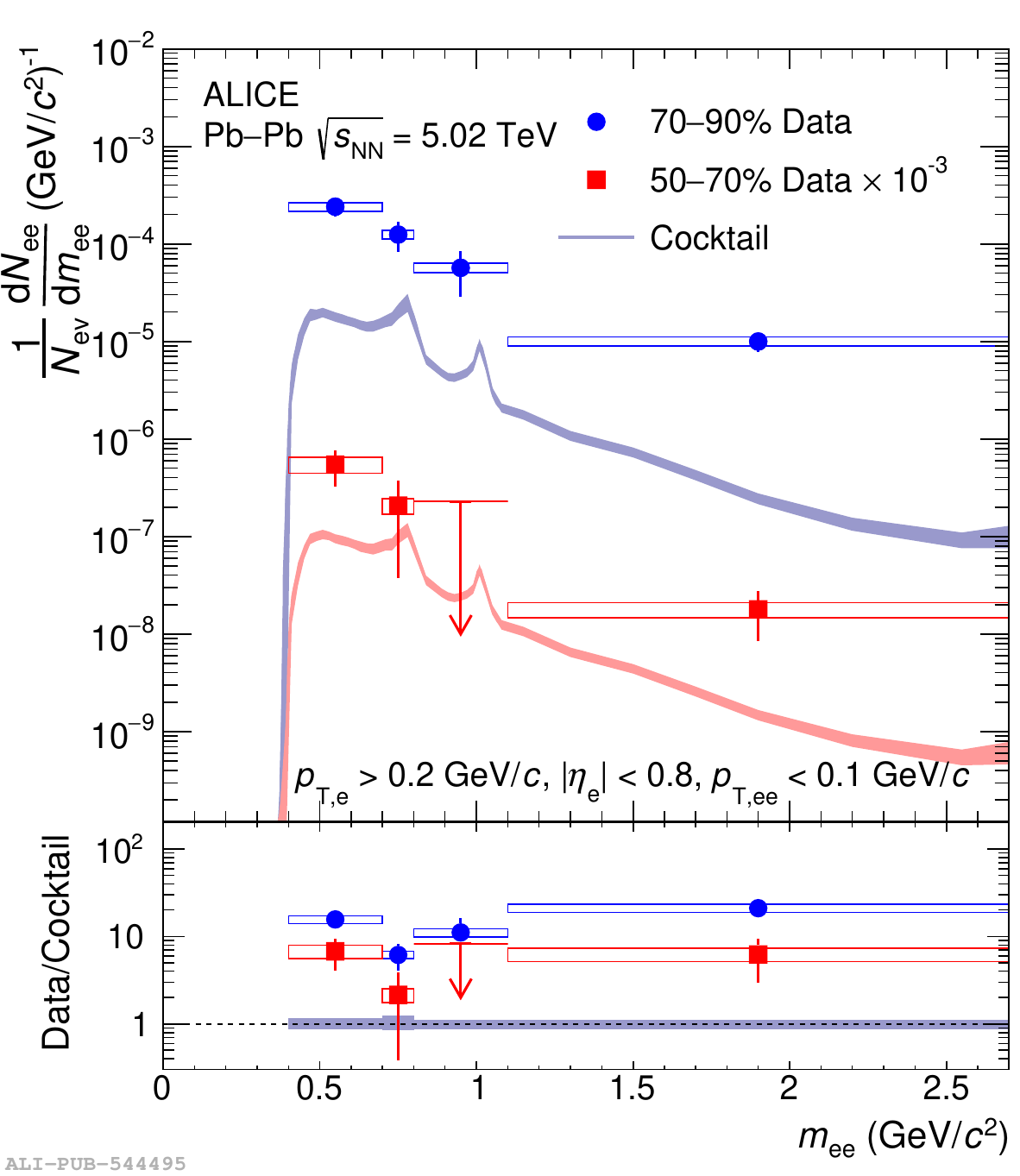}
    \caption{$e^+e^-$ invariant mass spectra in peripheral (70--90\%) and semi-peripheral (50--70\%) Pb--Pb collisions at $\sqrt{s_{NN}}=$ 5.02 TeV. Data are compared with the expected $e^+e^-$ from known hadronic contributions. Vertical bars and boxes represent the data statistic and systematic uncertainties, respectively. Bands show the uncertainties of the hadronic cocktail. Arrows indicate upper limits at 90\% confidence level.}
    \label{fig:mee_spectra}
\end{figure}

ALICE measured dielectron photoproduction with nuclear overlap at midrapidity within the ALICE central barrel acceptance ($|\eta_e| <$~0.8). The details of the analysis can be found in~\cite{ALICE:2022hvk}.
The efficiency-corrected $e^+e^-$ invariant mass spectra at low $\pt$, $p_{\rm T,ee} <$~0.1 GeV/$c$ in peripheral (70--90\% centrality class) and semi-peripheral (50--70\% centrality class) Pb--Pb collisions at $\sqrt{s_{NN}}=$ 5.02 TeV is shown in Fig.~\ref{fig:mee_spectra}.
Data are compared to the expected dielectron yield from known hadron decays produced in Pb--Pb collisions, called the hadronic cocktail. It is calculated with a fast simulation of the ALICE central barrel.
An excess compared to known hadronic expectation is observed for both centrality classes. A larger significance is found for the most peripheral Pb--Pb collisions (70--90\%) with a factor of about 20.

\begin{figure}
    \centering
    \includegraphics[scale=0.35]{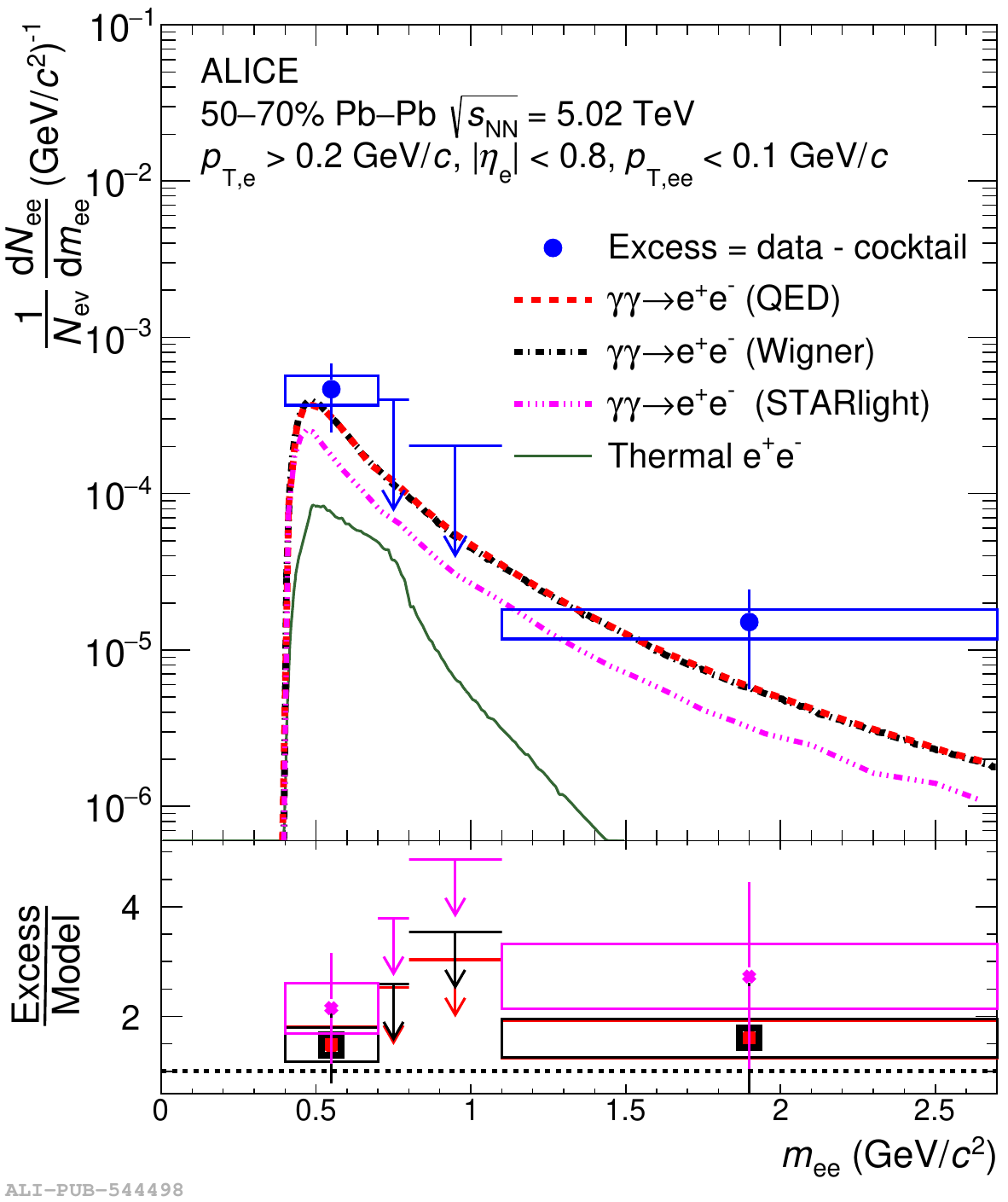}
    \includegraphics[scale=0.35]{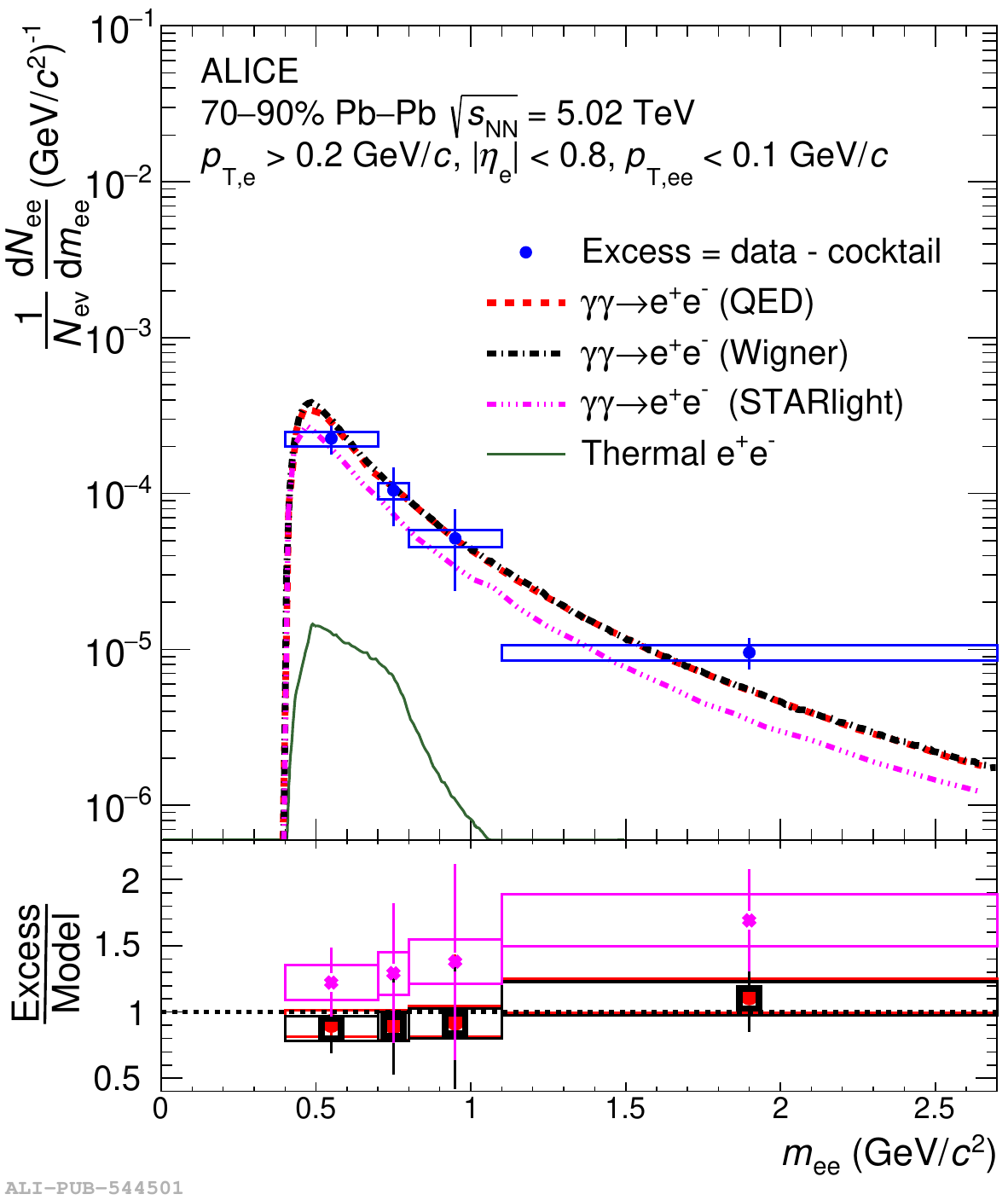}
    \caption{$e^+e^-$ invariant mass spectra of excess yield after subtraction of the hadronic cocktail to the data in semi-peripheral (left panel) and peripheral (right panel) Pb--Pb collisions at $\snn =$~5.02 TeV. The excess is compared with several calculations for photoproduction of electron pairs. Vertical bars and boxes represent the data statistic and systematic uncertainties, respectively. Arrows indicate upper limits at 90\% confidence level.} \label{fig:ee_cocktail_substraction_vs_mee}
\end{figure}

The $e^+e^-$ invariant mass distributions where the hadronic cocktail contribution is subtracted to the data are shown in Fig.~\ref{fig:ee_cocktail_substraction_vs_mee} for $p_{\rm T,ee} <$~0.1 GeV/$c$. The excess yield is compared with different calculations for dielectron photoproduction which are lowest-order QED calculations~\cite{Zha:2018tlq,Brandenburg:2021lnj}, Wigner formalism~\cite{Klusek-Gawenda:2020eja} and the STARlight MC generator using the equivalent photon approach~\cite{Klein:2016yzr,Klein:2018cjh}. On one hand, the STARlight model uses the $k_{\rm T}$-factorisation method, where the single-photon distribution is integrated over all transverse distances to obtain the shape of the $k_{\rm T}$--distribution. On the other hand the models based on QED calculations and Wigner formalism include the impact parameter dependence of the photon $k_{\rm T}$--distribution. The expected contributions from the thermal dielectrons, estimated with an expanding fireball model including in-medium broadened $\rho$ spectral function~\cite{Rapp:1999us,vanHees:2007th,Rapp:2013nxa} are also shown. The latter is expected to be at least one order of magnitude smaller than the measured excess.

\begin{figure}
  \begin{subfigure}{0.31\textwidth}
    \includegraphics[width=\linewidth]{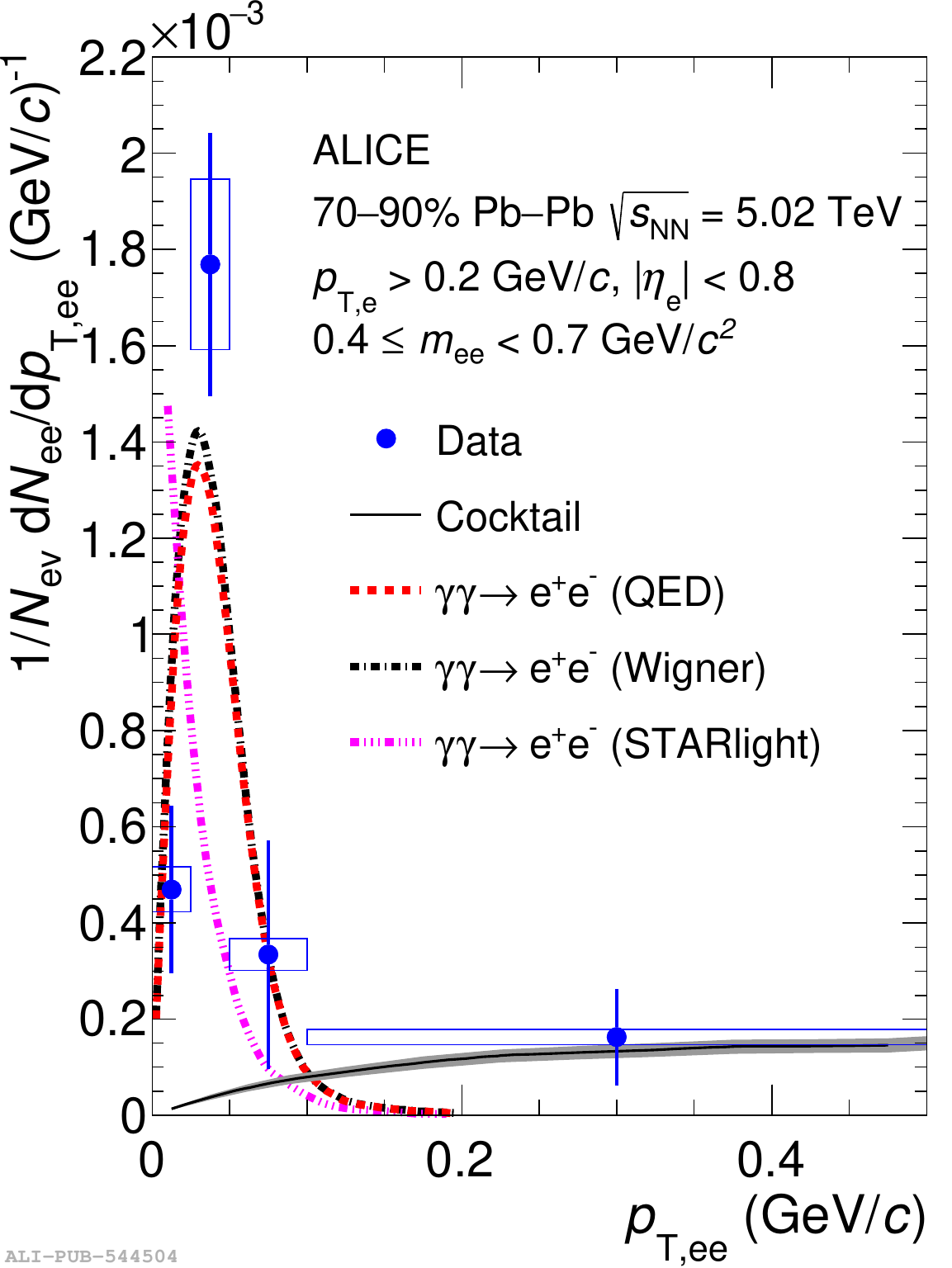}
    \caption{} \label{fig:ee_cocktail_vs_pT_a}
  \end{subfigure}%
  \hspace*{\fill}   
  \begin{subfigure}{0.31\textwidth}
    \includegraphics[width=\linewidth]{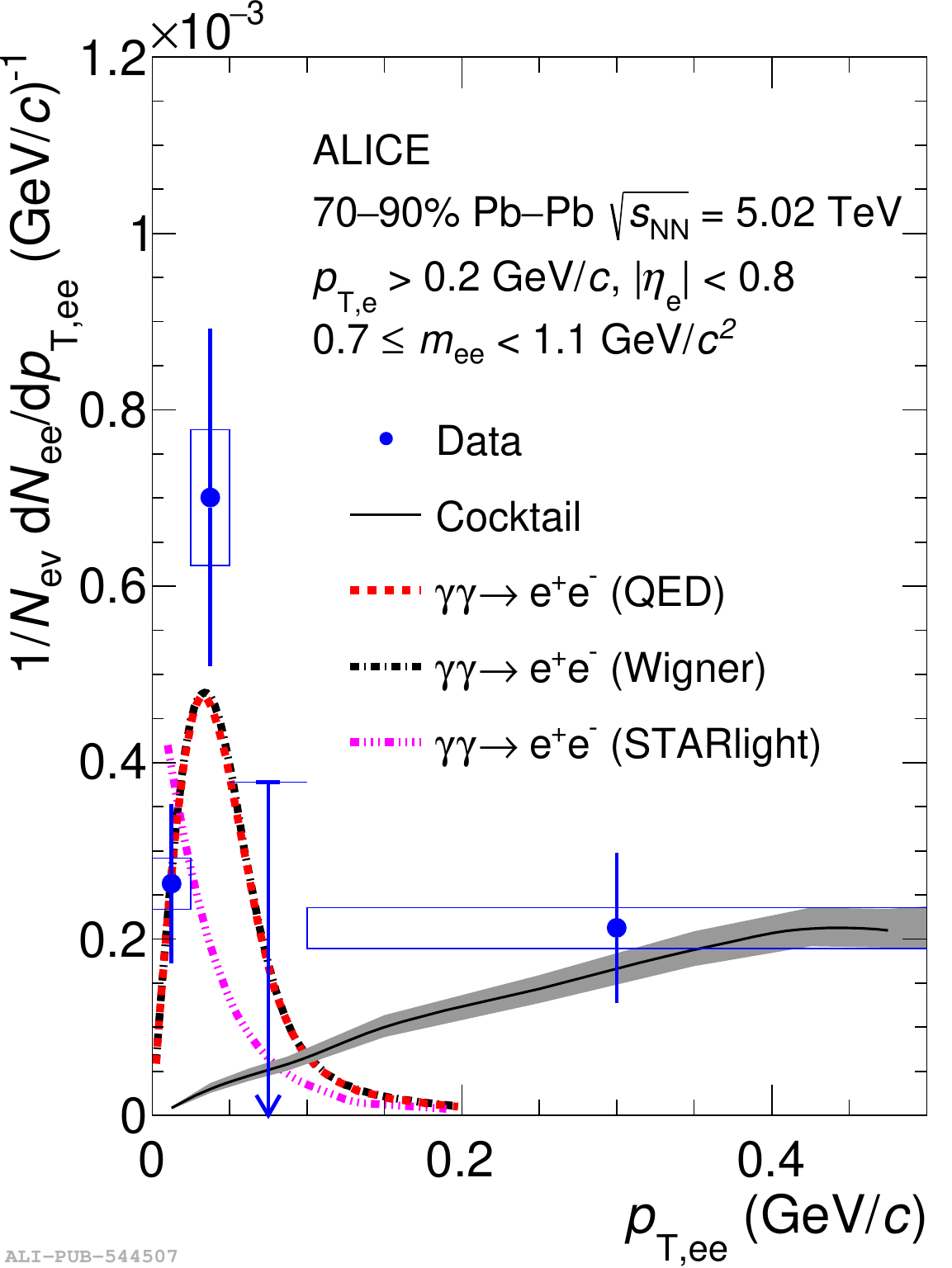}
    \caption{} \label{fig:ee_cocktail_vs_pT_b}
  \end{subfigure}%
  \hspace*{\fill}   
  \begin{subfigure}{0.31\textwidth}
    \includegraphics[width=\linewidth]{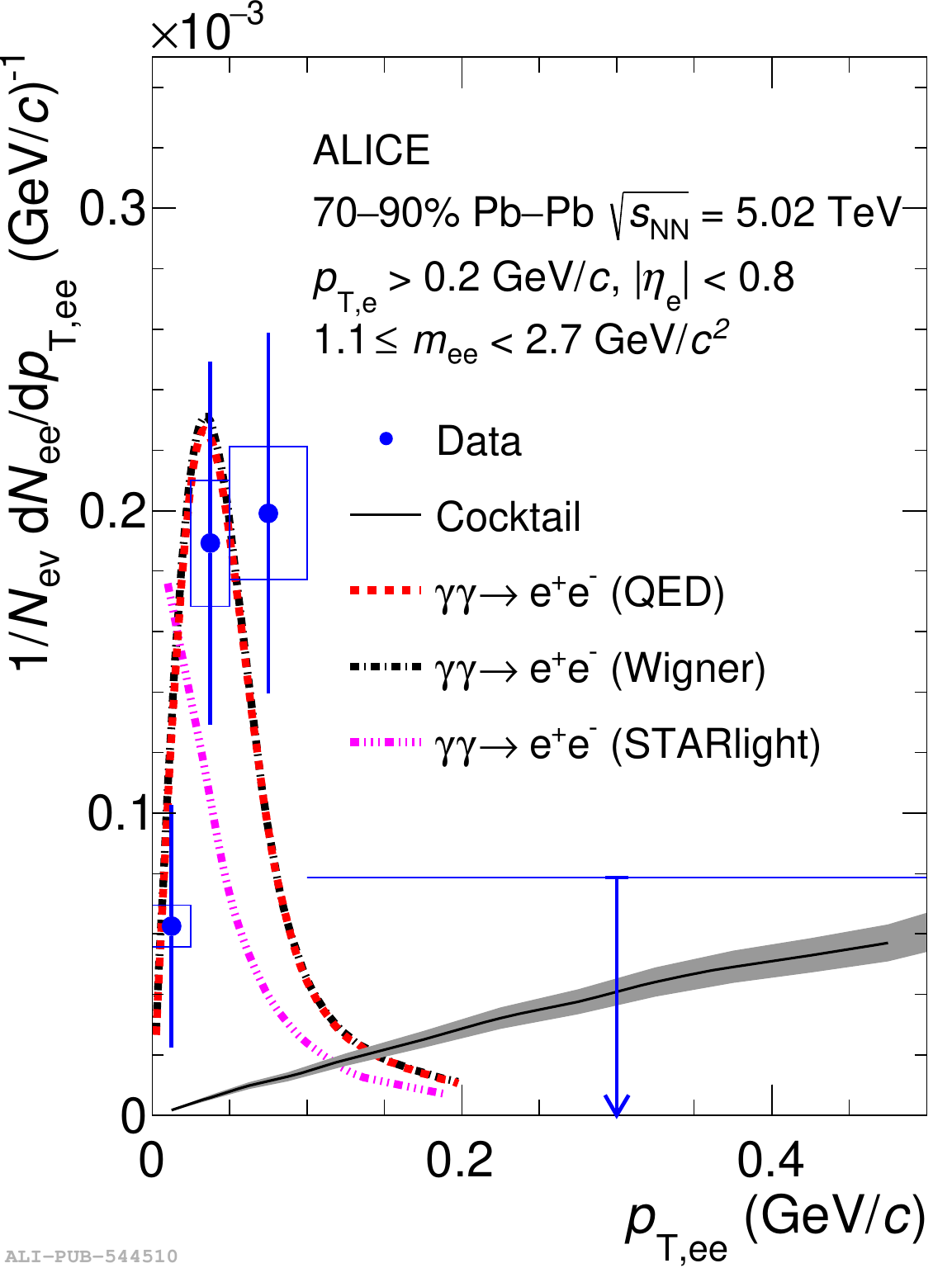}
    \caption{} \label{fig:ee_cocktail_vs_pT_c}
  \end{subfigure}
    \caption{$e^+e^-$ $\pt$ spectra. The data are compared with a cocktail of $e^+e^-$ coming from known hadronic contributions and with several calculations for photoproduction of electron pairs. Vertical bars and boxes represent the data statistic and systematic uncertainties, respectively. Arrows indicate upper limits at 90\% confidence level.} \label{fig:ee_cocktail_vs_pT}
\end{figure}

The inclusive dielectron $p_{\rm T,ee}$ spectra are shown for three different mass ranges in peripheral Pb--Pb collisions at $\sqrt{s_{NN}}=$ 5.02 TeV in Fig.~\ref{fig:ee_cocktail_vs_pT}. While the data are well described by the hadronic cocktail for higher $p_{\rm T,ee}~\geq$~0.1~GeV/$c$, a clear peak is observed for $p_{\rm T,ee}<$~0.1 GeV/$c$ in all $m_{ee}$ ranges. The latter is reproduced by the lowest-order QED calculations and the Wigner formalism, where both models include the impact parameter dependence of the photon $k_{\rm T}$--distribution. 
However, the STARlight model is rising down to $p_{\rm T,ee}=$ 0 and fails to reproduce the excess.

\FloatBarrier

\subsection{$\jpsi$ photoproduction measurements at forward rapidity}

\begin{figure}
    \centering
    \includegraphics[scale=0.425]{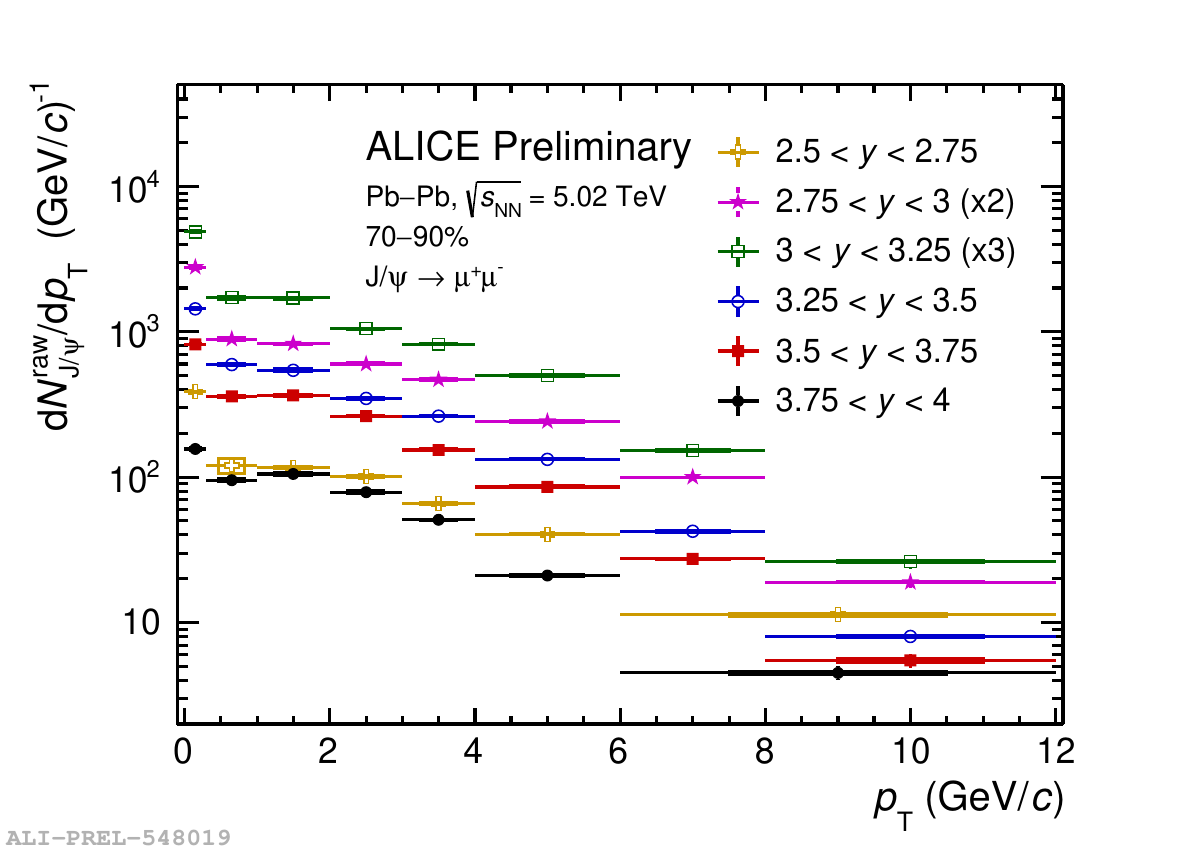}
    \includegraphics[scale=0.425]{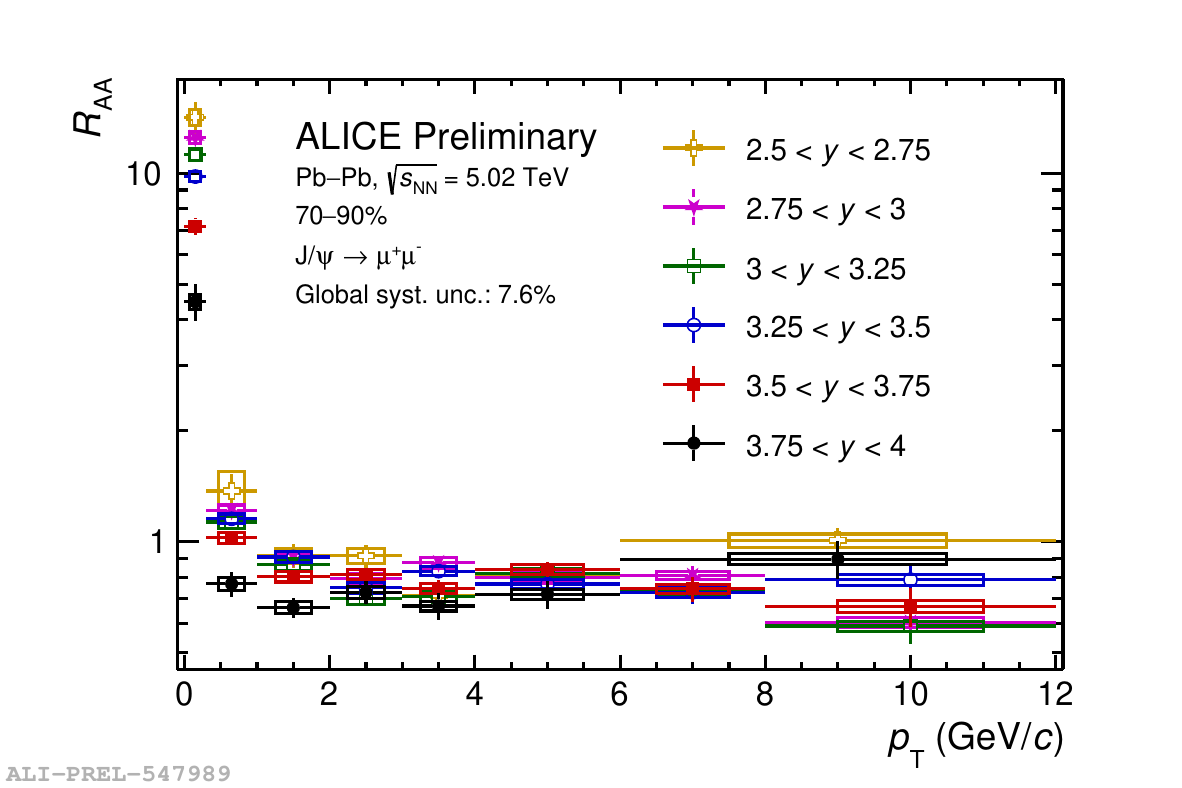}
    \caption{$\pt$--differential raw $\jpsi$ yield (left panel) and $\jpsi$ $R_{AA}$ (right panel) in six different rapidity intervals in Pb--Pb collisions at $\snn =$~5.02 TeV in the centrality class 70--90 \%. Vertical bars and boxes represent the data statistic and systematic uncertainties, respectively.}
    \label{fig:cohPhotoprodRawYieldAndRaaAsPt}
\end{figure}

The first measurement of the coherently photoproduced $\jpsi$ $y$--differential cross section in the dimuon decay channel is now described.
The $\jpsi$ raw yield is extracted at forward rapidity in Pb--Pb collisions at $\snn =$~5.02 TeV in the centrality class 70--90~\% in six different rapidity intervals within the muon spectrometer acceptance 2.5~$<y<$~4 as shown in the left panel of Fig.~\ref{fig:cohPhotoprodRawYieldAndRaaAsPt}. An excess is observed for $\pt <$~0.3 GeV/$c$ in all the rapidity intervals. The nuclear modification factor ($R_{AA}$), used to quantify nuclear effects in heavy-ion collisions with respect to pp collisions, is defined as
\begin{equation}
    R^{\jpsi}_{AA} = \frac{Y^{\jpsi}_{AA}}{\left<T_{AA}\right> \cdot \sigma^{\jpsi}_{pp}},
    \label{eq:RAA}
\end{equation}
where $Y^{\jpsi}_{AA}$ is the measured yield of $\jpsi$ in A--A collisions, $\left<T_{AA}\right>$ the average nuclear overlap function and $\sigma^{\jpsi}_{pp}$ the $\jpsi$ cross section in pp collisions. On the right panel of Fig.~\ref{fig:cohPhotoprodRawYieldAndRaaAsPt} the $R_{AA}$ of the $\jpsi$ is displayed. For all rapidity intervals, a strong increase of the $R_{AA}$ for $\pt<0.3$ GeV/$c$ is observed.

The excess yield is estimated by subtracting the hadronic $\jpsi$ yield to the raw $\jpsi$ yield.
The estimation of the hadronic yield is data driven and is performed using the $\jpsi$ $R_{AA}$, the $\jpsi$ cross section in pp collisions at the same center-of-mass energy and the hadronic $\jpsi$ Acceptance-Efficiency ($\mathcal{A}\times\mathcal{E}$) in Pb--Pb collisions.
The coherent $\jpsi$ yield is then obtained by correcting the excess from the fraction of incoherently photoproduced $\jpsi$ ($f_I$) and the fraction of coherent $\jpsi$ from coherently photoproduced $\psi$(2S) feed-down ($f_D$). Both $f_I$ and $f_D$ are taken from UPC measurements~\cite{ALICE:2019tqa}. The cross section of the coherently photoproduced $\jpsi$ is described as
\begin{equation}
    \frac{\mathrm{d}\sigma^{\jpsi}_{coh}}{\mathrm{d}y} = \frac{Y^{\jpsi}_{coh}}{\mathrm{BR}(\jpsi \rightarrow \mu^+\mu^-) \times (\mathcal{A}\times\mathcal{E})^{coh \jpsi}_{AA} \times \mathcal{L}_{int} \times \Delta y},
    \label{eq:cohCS}
\end{equation}
where $Y^{\jpsi}_{coh}$ is the obtained coherently photoproduced $\jpsi$ yield, $\mathrm{BR}(\jpsi \rightarrow \mu^+\mu^-)$ is the branching ratio for the $\jpsi$ to dimuon decay with $\mathrm{BR}=$ 5.961 $\pm$ 0.033~\%, $(\mathcal{A}\times\mathcal{E})^{coh \jpsi}_{AA}$ is the Acceptance-Efficiency of the coherently photoproduced $\jpsi$ in Pb--Pb collisions and $\mathcal{L}_{int}=$ 755.7 $\pm$ 18.9 $\mu$b$^{-1}$ is the integrated luminosity on the Pb--Pb data sample used.

\begin{figure}
    \centering
    \includegraphics[scale=0.4]{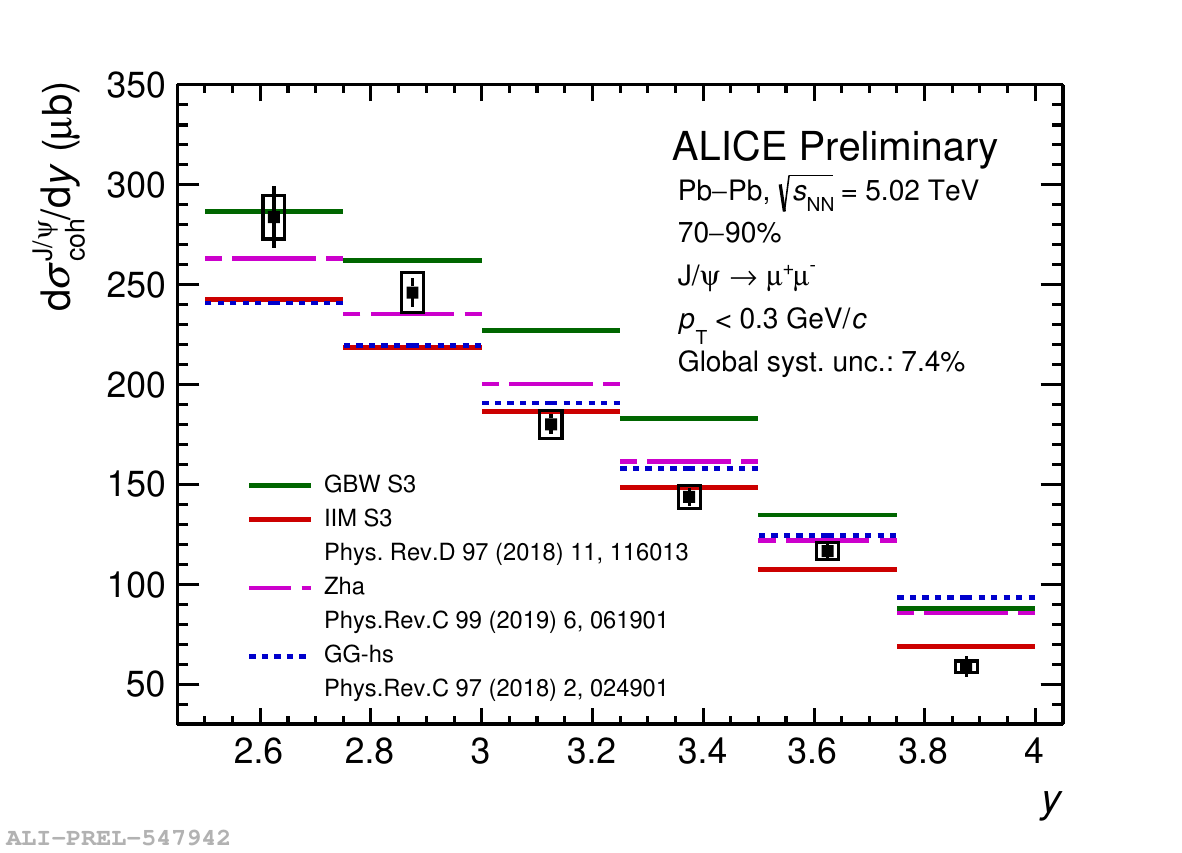}
    \includegraphics[scale=0.4]{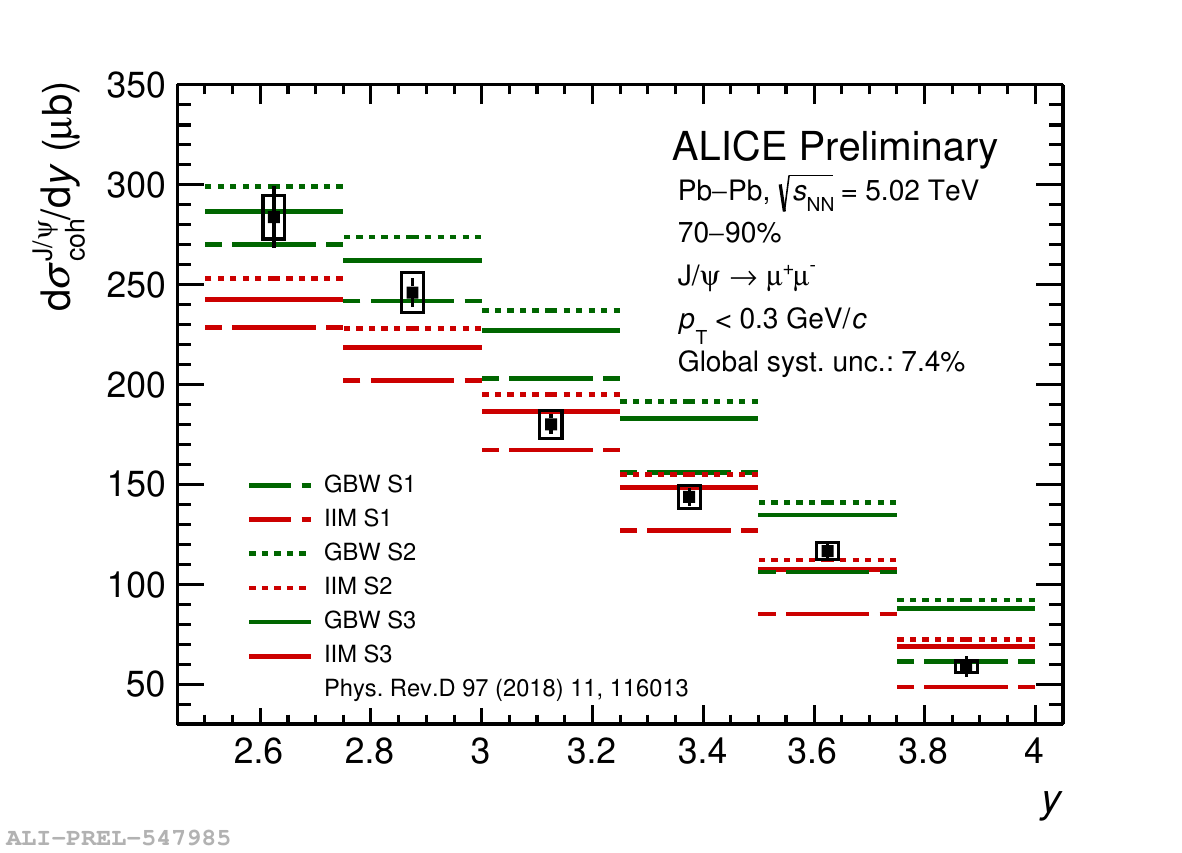}
    \caption{Coherently photoproduced $\jpsi$ cross section as a function of rapidity in Pb--Pb collisions at $\snn =$~5.02~TeV in the centrality class 70--90 \%. Data are compared to several UPC models adapted for Pb--Pb collisions with nuclear overlap. Vertical bars and boxes represent the data statistic and systematic uncertainties, respectively.}
    \label{fig:cohSigmaAsY}
\end{figure}

The coherently photoproduced $\jpsi$ $y$-differential cross section is shown in Fig.~\ref{fig:cohSigmaAsY}. The measurement shows a strong rapidity dependence and is compared to several models of $\jpsi$ photoproduction in UPC, adapted to take into account the nuclear overlap. 
The hot-spot model (GG-hs)~\cite{Cepila:2017nef} proposes that the photon flux is constrained by the impact parameter range for a given class of centrality.
In the model suggested by Zha~\cite{Zha:2018jin}, an emitted photon from one nucleus interacts with a pomeron emitted from the spectator region of the other one.
The GBW/IIM~\cite{GayDucati:2018who} models offer three scenarios. The first scenario has no relevant modifications with respect to UPC calculations. In the second one, only the photon reaching the spectator nucleon region is considered and the photonuclear cross section remains unmodified.
The third scenario includes the same effective photon flux as the second one but the nuclear overlap region is excluded from the calculation of the photonuclear cross section. The three scenarios are displayed on the right panel of Fig.~\ref{fig:cohSigmaAsY}. They are all able to qualitatively describe the cross section.

\begin{figure}
    \centering
    \includegraphics[scale=0.3]{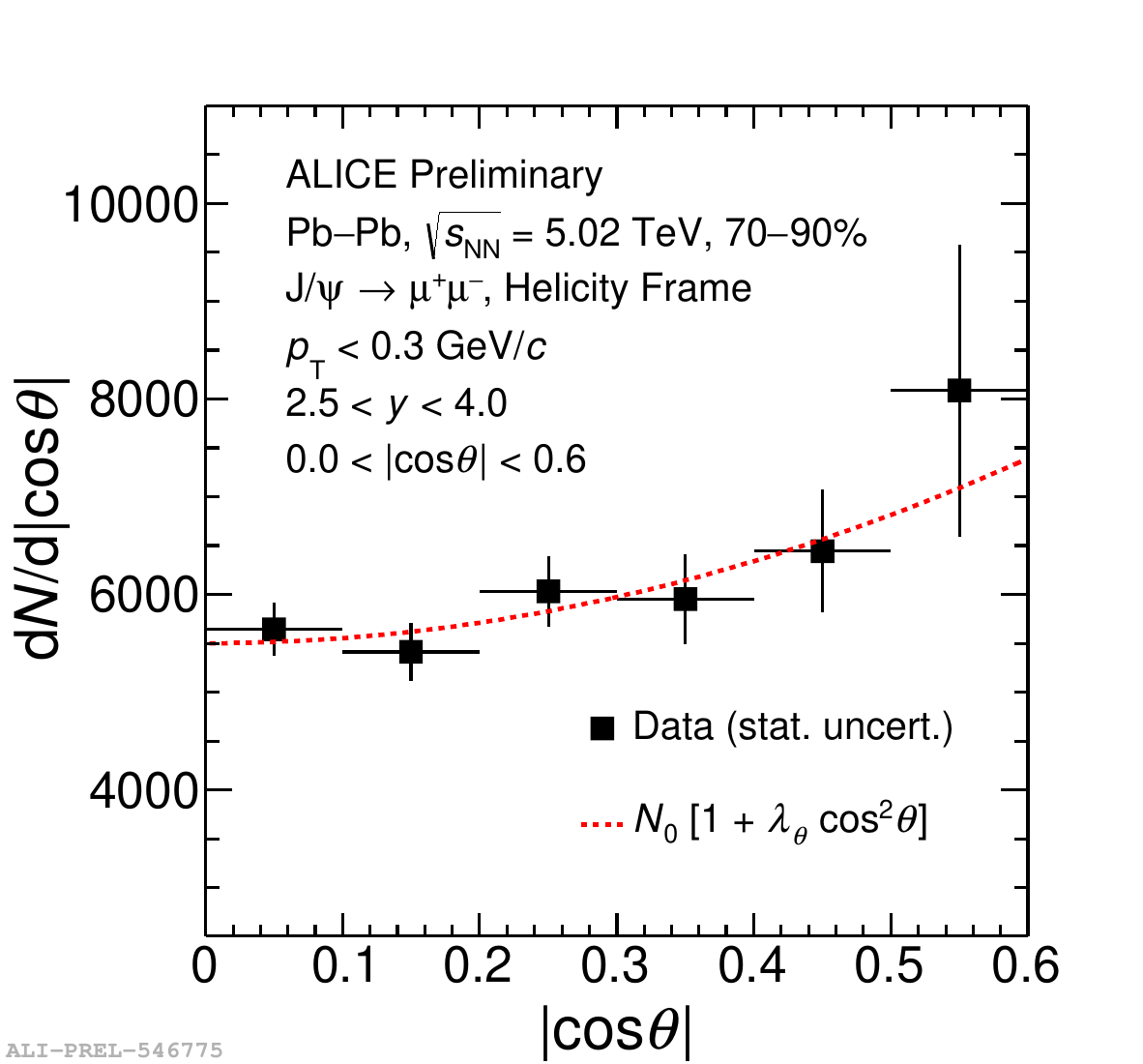}
    \includegraphics[scale=0.3]{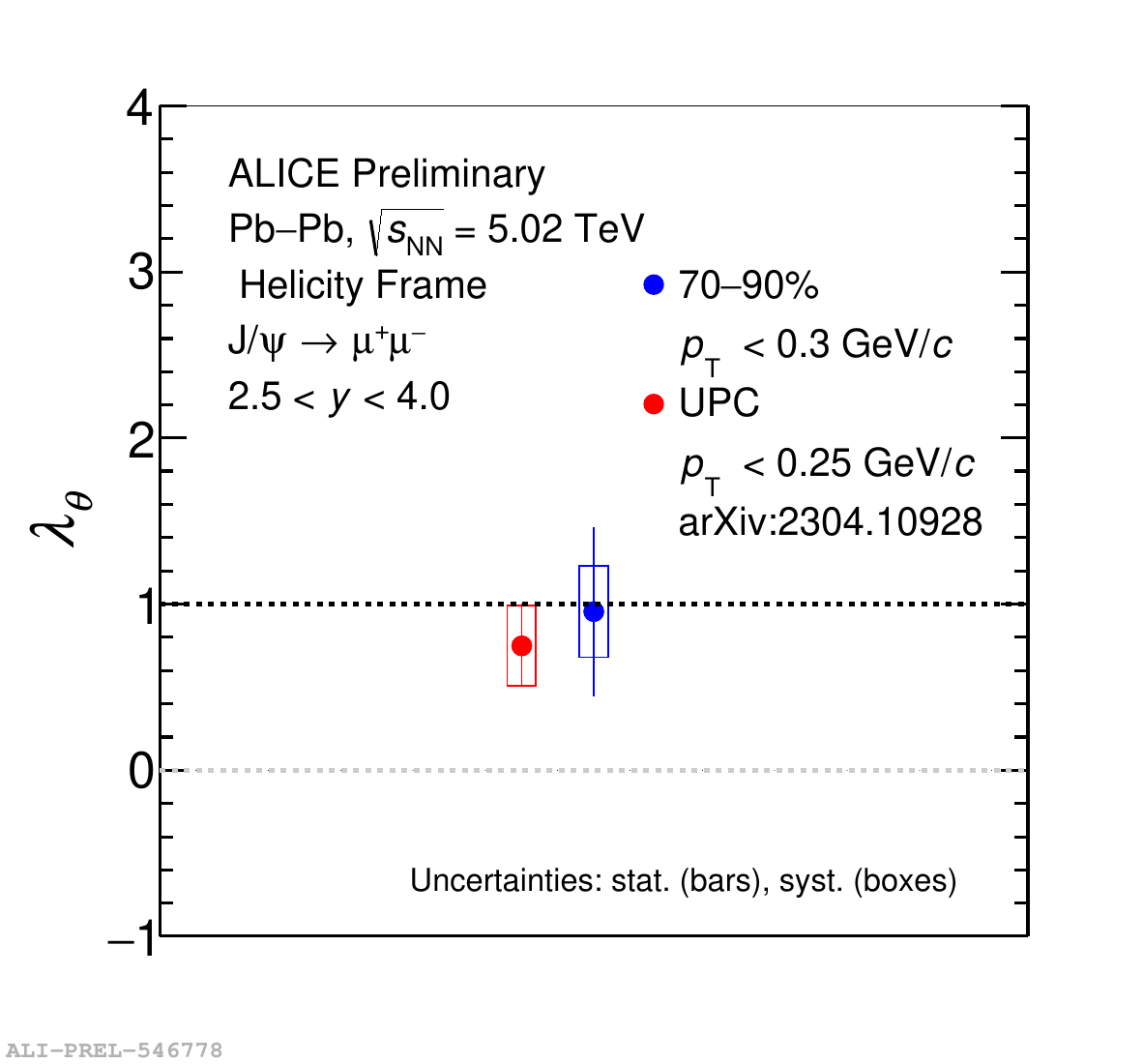}
    \caption{Dimuon angular distribution in semi-peripheral (left panel) and peripheral (right panel) Pb--Pb collisions at $\snn =$~5.02 TeV and $\lambda_\theta$ value compared to UPC measurements~\cite{ALICE:2023svb}. Vertical bars and boxes represent the data statistic and systematic uncertainties, respectively.}
    \label{fig:jpsipolarization}
\end{figure}

The last ALICE measurement reported in these proceedings is the inclusive $\jpsi$ polarization measured at forward rapidity, 2.5~$<y<$~4, in Pb--Pb collisions at $\snn =$~5.02 TeV in the 70--90\% centrality class.
The $\jpsi$ signal is extracted in the dimuon decay channel in six different $\cos\theta$ intervals at low $\pt$, $\pt<$~0.3 GeV/$c$ as shown in the left panel of Fig.~\ref{fig:jpsipolarization}.
$\theta$ is the angle between the muon direction in the $\jpsi$ rest frame and the flight direction of the $\jpsi$ in the Pb--Pb center-of-mass frame.
The $\cos \theta$ distribution is fitted with the dimuon angular distribution expressed as 
\begin{equation}
    W(\cos\theta,\phi )
    \propto 
    \frac{1}{3+\lambda_\theta}
    \left[1 + \lambda_\theta \cos^2\theta + \lambda_\phi \sin^2 \theta \cos2\phi + \lambda_{\theta\phi} \sin2\theta \cos\phi\right],
    \label{eq:angulardistribution}
\end{equation}
where $\lambda_\phi$ and $\lambda_{\theta\phi}$ are set to zero. The polarization of the $\jpsi$ is transverse when $\lambda_\theta =$ 1 and longitudinal when $\lambda_\theta =$ -1. A $\lambda_\theta$ value equal to zero indicates no polarization.
The measured angular distribution suggests a transverse polarization of the inclusive $\jpsi$ for $\pt<$~0.3~GeV/$c$ as expected for photoproduced $\jpsi$ and from s--channel helicity conservation. As shown on the right panel of Fig.~\ref{fig:jpsipolarization}, the $\lambda_\theta$ value obtained with the present measurement is close to unity and compatible with the one obtained for coherently photoproduced $\jpsi$ as described in~\cite{ALICE:2023svb}.

\section{Conclusion}
An excess of dielectron photoproduction measurement at low $\pt$ in the 50--70\% and 70--90\% centrality classes in Pb--Pb collisions at $\snn=$~5.02 TeV was presented. The two-photons models that include the impact parameter dependence of the photon $k_{\rm T}$--distribution are able to describe the data.
In addition, the first $y$--differential cross section of the coherent $\jpsi$ photoproduction and the first inclusive $\jpsi$ polarization measurements for $\pt<$~0.3~GeV/$c$ in the 70--90\% centrality class in Pb--Pb collisions at $\snn=$~5.02 TeV have been discussed.
Several models developed for UPC and adapted to consider the nuclear overlap are able to qualitatively reproduce the $y$--differential cross section. The inclusive $\jpsi$ polarization measurement is consistent with a transverse polarization as expected for photoproduced $\jpsi$ and from the s-channel helicity conservation. The measured polarization is in agreement with previous measurements in UPC.
The ALICE Run 3 and Run 4 will provide a larger Pb--Pb data sample. The increased statistics will allow one to study the dileptons and $\jpsi$ photoproduction in more central events both at mid and forward rapidities. A better precision on the cross section and the polarization measurements is also expected. Such improvement will be useful to provide further constraints on the theoretical models. Finally, this will help to access other coherently photoproduced excited states like the $\psi$(2S) and to look for possible final-state medium effects on the photoproduced vector mesons.

\section*{Acknowledgments}

I would like to thanks the organizers for giving me the opportunity to present ALICE collaboration results at the first edition of the UPC workshop that took place in such a marvelous place as Playa Del Carmen, Mexico.

\printbibliography

\end{document}